*Article*

# CMOS-Compatible Ultrathin Superconducting NbN Thin Films Deposited by Reactive Ion Sputtering on 300 mm Si Wafer

Zihao Yang [1], Xiucheng Wei [2], Pinku Roy [3], Di Zhang [4], Ping Lu [5], Samyak Dhole [3], Haiyan Wang [4], Nicholas Cucciniello [3], Nag Patibandla [1], Zhebo Chen [1], Hao Zeng [2], Quanxi Jia [3,\*] and Mingwei Zhu [1,\*]

[1] Applied Materials Inc., Santa Clara, CA 95054, USA; zihao_yang@amat.com (Z.Y.); nag_patibandla@amat.com (N.P.); zhebo_chen@amat.com (Z.C.)
[2] Department of Physics, University at Buffalo—The State University of New York, Buffalo, NY 14260, USA; xiucheng@buffalo.edu (X.W.); haozeng@buffalo.edu (H.Z.)
[3] Department of Materials Design and Innovation, University at Buffalo—The State University of New York, Buffalo, NY 14260, USA; pinkuroy@buffalo.edu (P.R.); sdhole@buffalo.edu (S.D.); ngcuccin@buffalo.edu (N.C.)
[4] School of Materials Engineering, Purdue University, West Lafayette, IN 47907, USA; dizhang@lanl.gov (D.Z.); hwang00@purdue.edu (H.W.)
[5] Sandia National Laboratories, Albuquerque, NM 87185, USA; plu@sandia.gov
\* Correspondence: qxjia@buffalo.edu (Q.J.); mingwei_zhu@amat.com (M.Z.)

**Abstract:** We report a milestone in achieving large-scale, ultrathin (~5 nm) superconducting NbN thin films on 300 mm Si wafers using a high-volume manufacturing (HVM) industrial physical vapor deposition (PVD) system. The NbN thin films possess remarkable structural uniformity and consistently high superconducting quality across the entire 300 mm Si wafer, by incorporating an AlN buffer layer. High-resolution X-ray diffraction and transmission electron microscopy analyses unveiled enhanced crystallinity of (111)-oriented δ-phase NbN with the AlN buffer layer. Notably, NbN films deposited on AlN-buffered Si substrates exhibited a significantly elevated superconducting critical temperature (~2 K higher for the 10 nm NbN) and a higher upper critical magnetic field or $H_{c2}$ (34.06 T boost in $H_{c2}$ for the 50 nm NbN) in comparison with those without AlN. These findings present a promising pathway for the integration of quantum-grade superconducting NbN films with the existing 300 mm CMOS Si platform for quantum information applications.

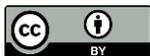

## 1. Introduction

Quantum information science has recently received extensive research attention due to its potential application in next-generation quantum computing [1,2], quantum sensing [3,4], and quantum communication [5,6]. Many of these quantum devices rely on high-quality superconducting thin films to generate, manipulate, and detect external stimuli such as photons, current, magnetic fields, heat, and others. Among the various superconducting materials, NbN stands out as one of the most widely used superconductors for a broad range of quantum devices. For example, superconducting nanowire single-photon detectors (SNSPDs) based on NbN and $Nb_xTi_{1-x}N$ have been demonstrated with superb ability in detecting single photons with high efficiency, low timing jitter, and low dark count rate [7–10]. A superconducting cavity electro-optic transducer based on NbN and $LiNbO_3$ enables bidirectional conversion between C-band and microwave photons with high internal efficiency [11]. Furthermore, superconducting

Josephson junctions based on epitaxial NbN and AlN provide a promising pathway for fabricating high-quality flux-bias-free flux qubits [12].

Among these NbN-based quantum devices, several are transitioning from early-stage proof-of-concept to more mature technologies, including multiple-pixel SNSPD arrays, superconducting qubit arrays, etc. [13–15]. Recently, there has been a growing interest in leveraging the 300 mm (12 inch) advanced CMOS fabrication environment for scaling up quantum devices due to the availability of state-of-the-art semiconductor manufacturing tools and high-precision process control [16–19]. Giewont et al. proposed the development of a large-scale photonic quantum chip using advanced CMOS and silicon-on-insulator (SOI) technologies, in which NbN SNSPDs could be integrated with other photonic components, e.g., waveguides, couplers, phase shifters, etc., for photonic quantum computing [18]. Wan et al. demonstrated a CMOS-compatible fabrication process for highly reliable Nb/Al–AlO$_x$/Nb trilayer stacks Josephson junctions free of error-prone lift-off process for superconducting quantum computing [19]. One of the most crucial steps in fulfilling this promise is achieving high quality and high uniformity superconducting films such as NbN on large-scale wafers. It is well known that the performance of NbN-based quantum devices is very sensitive to their crystallinity and superconducting properties [20,21]. Previously, large-scale high-quality NbN has only been demonstrated on an 8 inch (200 mm) Si wafer, with a focus on SNSPD device performance [17]. There is still a lack of fundamental study on the structural and superconducting properties of NbN thin films deposited on a full 300 mm Si wafer.

In this paper, we demonstrated, for the first time, high-quality NbN films deposited on an AlN buffered 300 mm Si wafer by the CMOS back-end of the line (BEOL) compatible HVM-ready system. The properties of the blanket NbN films, including thickness, sheet resistance, superconducting transition temperature, and others, were characterized at different locations on the 300 mm wafer level to assess their uniformity. Detailed microstructural properties were studied by high-resolution X-ray diffractometry (XRD), transmission electron microscopy (TEM), high-resolution scanning transmission electron microscopy (STEM), and energy-dispersive X-ray spectroscopy (EDX). Superconducting properties were studied by measuring the zero field and magnetic field-dependent superconducting transition temperatures, from which the upper critical fields were extracted.

## 2. Materials and Methods

Two sets of NbN samples with different thicknesses were deposited on Si wafers with and without an AlN buffer layer using Applied Materials 300 mm Endura® (Santa Clara, California, USA) high vacuum production system. The sample fabrication consisted of substrate surface treatment, AlN buffer layer deposition, and NbN layer deposition. All these steps were achieved within the Endura® vacuum system in a low base pressure environment without breaking the vacuum enabled by the multiple process chambers. In addition, the staged vacuum architecture and the high precision fast robotics enabled the HVM process of NbN thin films with high throughput. The AlN and NbN thin films were deposited via magnetron reactive sputtering on mechanical grade 300 mm (100) oriented Si wafers in Impulse® PVD process chambers that were mounted on the Endura® mainframe. Specifically, the AlN and NbN were deposited with metallic Al and Nb targets in an Ar/N$_2$ ambient at a substrate temperature of 400 °C. The deposition power and pressure were determined to achieve a moderate deposition rate of ~0.50 nm/s for AlN and ~0.55 nm/s for NbN, respectively. The N$_2$ to Ar gas flow ratio was carefully chosen to ensure stable nitride film deposition and to achieve uniform thin films with high structural and superconducting qualities on the entire wafer. The Si wafers were degassed at a temperature of 400 °C prior to the AlN buffer layer deposition. The AlN buffer layer thickness was fixed at 20 nm while the NbN layer thickness was varied at 5, 8, 10, 20, 30, and 50 nm. The NbN film thickness and sheet resistance at the wafer level were characterized by a Rigaku MFM310 X-ray fluorescence (XRF) (Shibuya-ku, Tokyo, Japan) system and a KLA RS-200 (Milpitas, California, USA) four-point probe station, respectively.

## 3. Results and Discussions

### 3.1. Full Wafer Analysis

The thickness and sheet resistance profile show superior uniformity ($1\sigma < 2.0\%$ and $1\sigma < 3.0\%$, respectively) for NbN films with different thicknesses. Representative thickness (d) and sheet resistance ($R_s$) mapping and the sample image taken on 20 nm NbN on a 300 mm Si wafer are shown in Figure 1a–c, respectively. The stability and reliability of the properties of these films were confirmed by the industrial standard extended run. In addition to thickness and resistance, the root mean square surface roughness of the NbN films analyzed by atomic force microscopy in a scan area of 5 μm × 5 μm was 0.6 nm and 3.2 nm for NbN films with a thickness of 5 nm and 50 nm, respectively. The structural and superconducting property measurements were carried out at different locations on 50 nm NbN deposited on an AlN-buffered 300 mm Si wafer. As summarized in Table 1, the critical temperature ($T_C$), superconducting transition width ($\Delta T_C$), full-width at half-maximum (FWHM) of the NbN (111) X-ray diffraction peak, and the grain size exhibited narrow distributions. One thing to notice is that the $T_C$ values presented in Table 1 are slightly higher than those shown in Figure 4 measured at a later time, possibly due to the queue time effect on these samples. Detailed analyses of the XRD and transport measurement results are discussed later. Given these characterization results, these NbN films (both with and without an AlN buffer layer) exhibited a high level of homogeneity in terms of structural and superconducting properties across the entire 300 mm wafer.

**Table 1.** Structural and superconducting properties of 50 nm NbN on an AlN-buffered 300 mm Si wafer.

| Location | $T_C$ (K) | $\Delta T_C$ (K) | (111) FWHM (°) * | Grain Size (nm) |
|---|---|---|---|---|
| R = 0 mm | 16.66 | 0.104 | 0.224 | 38.9 |
| R ~ 72.5 mm | 16.62 | 0.105 | 0.237 | 36.8 |
| R ~ 145 mm | 16.41 | 0.075 | 0.266 | 32.8 |

* FWHM of the NbN (111) diffraction peak in 2θ-ω scan.

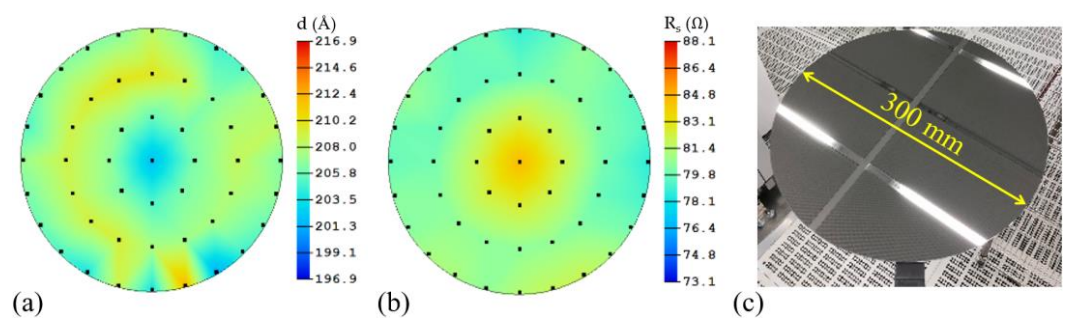

**Figure 1.** Surface mapping of 20 nm NbN films on a 300 mm Si wafer: (**a**) thickness, (**b**) sheet resistance, and (**c**) sample image.

### 3.2. Structural Properties

XRD measurements were carried out using a Malvern Panalytical Empyrean system (Malvern, United Kingdom). Figure 2 shows the 2θ-ω scan of NbN films deposited on Si with and without a 20 nm AlN buffer layer, with thicknesses ranging from 5 to 50 nm. Given the appearance of a shoulder-like diffraction peak around 2θ angles from 35° to 36.5° for the NbN films on AlN buffered Si, we used the pseudo-Voigt function to fit the diffraction peaks. The inset in Figure 2 shows an example of the fitting of the 30 nm NbN with a 20 nm AlN buffer layer. Two predominant diffraction peaks corresponding to δ-NbN (111) and AlN (0002) were observed for NbN films with an AlN buffer layer. Similar results have been reported by other research groups [22,23]. It is noted that the weak peak around 33° was from Si (200) reflection, where the intensity and the shape can vary, depending on the alignment of the phi-angle of the four-circle X-ray diffractometer. The preferential out-of-plane c-axis orientation of the AlN buffer layer was a result of its columnar growth mode evidenced by the TEM images shown in Figure 3. It is also interesting to note that while the AlN (0002) diffraction peak position was located at 36.01 ± 0.02°, the NbN (111) diffraction peak position shifted from 35.25° for the thinnest NbN film to a slightly larger value of 35.43° for the thickest NbN film. This implies the presence of interfacial strain due to the lattice mismatch between AlN and NbN. As shown in Figure 3, the columnar size increased with the NbN film thickness, e.g., about 9 nm and 20 nm for NbN films with a thickness of 5 nm and 50 nm, respectively. As the NbN films on top of the (0002)-oriented AlN buffer layer showed a high degree of preferential orientation along the [111] direction, lattice strains resulting from the growth of preferentially oriented grains could introduce measurable peak shift from the X-ray diffraction. The selected area electron diffraction (SAED) patterns from the cross-sectional TEM analysis (as shown in Figure 3) revealed epitaxial-like features. In other words, preferentially oriented NbN and AlN layers were formed, in spite of the use of BEOL-compatible deposition temperature at 400 °C. This was quite remarkable considering that high-quality epitaxial NbN films typically require a much higher growth temperature (~600 °C or above) via sputtering [23,24]. As for the NbN deposited on Si without an AlN buffer layer, no distinct diffraction peaks were observed until the NbN film thickness reached 50 nm, indicating lower crystallinity of the thinner NbN films.

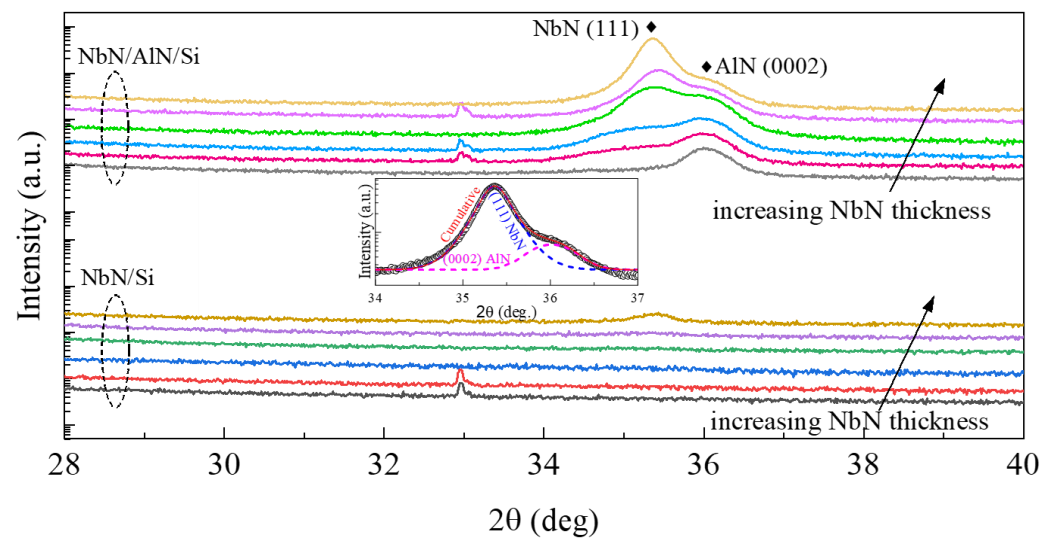

**Figure 2.** X-ray 2θ-ω scan for NbN films with different thicknesses (5, 8, 10, 20, 30, and 50 nm) with and without AlN buffer layer. The arrows shown in the figure indicate the increase in NbN film thickness from 5 nm to 50 nm, where the spectra are color-coded and vertically shifted for clarity. The inset shows the pseudo-Voigt fitting of the NbN and AlN diffraction peaks.

Detailed microstructural analysis, including TEM, STEM, and EDX, was conducted using the Thermo Fisher Scientific Talos F200X system (Waltham, Massachusetts, USA) operated at 200 kV and an aberration-corrected FEI Titan microscope equipped with a high-brightness Schottky field emission electron source operated at 300 kV. The cross-sectional TEM samples were prepared by the traditional method, including grinding, polishing, dimpling, and a final ion milling step (PIPS 695 precision ion polishing system, Gatan Inc., Pleasanton, California, USA). The TEM images of a 50 nm thick NbN film deposited on the AlN buffer layer on Si are shown in Figure 3a, where the inset shows the SAED pattern taken along the Si <011> zone axis [25]. Columnar and highly textured growth of NbN and AlN can clearly be observed. Small NbN grains can be observed in the bottom ~25 nm thickness, whereas larger columnar grains are found in regions above 25 nm thickness. This observation can be used to explain the shift of the NbN (111) diffraction peak resulting from different film thicknesses shown in Figure 2. The SAED pattern confirms the highly textured growth of NbN evidenced by the slightly arced diffraction dots. Figure 3b–d shows STEM images taken under the high-angle annular dark field mode (STEM-HAADF) and the corresponding EDX elemental maps of the film [25]. The STEM-HAADF image shows evident columnar growth of NbN film, and the EDX maps demonstrate the sharp interfaces between the individual layers. The high-resolution STEM images of a 5 nm thick NbN film deposited on the AlN buffered Si are shown in Figure 3e–g. The fast Fourier transform (FFT) patterns obtained from the top layer NbN, buffer layer AlN, AlN/Si interface, and Si substrate are all shown as insets in Figure 3f,g. It can be seen that the wurtzite AlN buffer layer shows a [0001] growth direction on Si, while the top NbN film presents a (111) plane growth. The observed growth mode and crystal orientations of AlN and NbN agree well with the XRD data discussed in Figure 2.

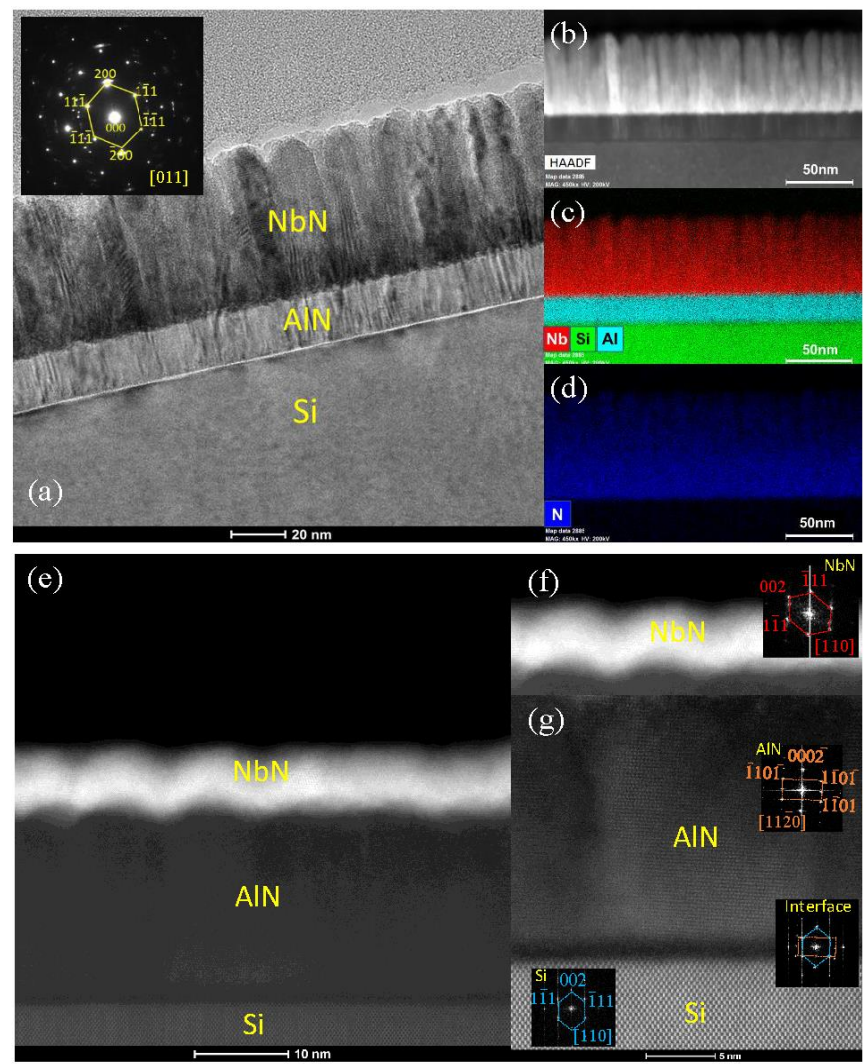

**Figure 3.** TEM images of: (**a**) 50 nm NbN on 20 nm AlN buffered Si, the inset shows the SAED pattern taken along the Si <011> zone axis of the heterostructure; (**b**) STEM-HAADF image; (**c**) and (**d**) EDX maps for Nb, Si, Al, and N taken from the region in (**b**) [25]; (**e**) STEM-HAADF of 5 nm NbN on 20 nm AlN buffered Si; (**f**,**g**) high-resolution STEM-HAADF images with FFT patterns taken from Si, Si/AlN interface, AlN, and NbN regions.

*3.3. Superconducting Properties*

Superconducting property measurements were carried out in a Quantum Design Physical Property Measurement System (PPMS). The temperature-dependent DC resistivity of the NbN films was measured in a four-probe configuration. Figure 4a shows the thickness dependence of the extracted superconducting transition temperature and transition width of the NbN films on Si with (open symbols) and without (solid symbols) an AlN buffer layer. The transition width is defined as the temperature difference between 90% and 10% of the normal resistance, while the transition temperature is defined as the temperature at 50% of the normal resistance. The transition width was observed to decrease monotonically from ~1.2 K to ~0.2 K with increasing NbN film thickness from 5 nm to 50 nm, for both samples with and without the AlN buffer layer. This trend directly reflected a change in the overall NbN film homogeneity and the crystallinity with different NbN thicknesses. In addition, the relatively narrower transition width of NbN films with an AlN buffer layer, compared with those without the NbN buffer layer, indicated enhanced film homogeneity. The transition temperature, on the other hand, increased as film thickness increased. For instance, the $T_c$ varied from 9.9 K to 14.5 K for films without the AlN buffer layer, and from 11.5 K to 15.6 K for films with an AlN buffer layer. These $T_c$ values were higher compared with the previous reports on NbN films deposited on glass or GaAs substrates via sputtering, but lower than the epitaxial NbN grown on AlN/sapphire via MOVPE and sputtering, which implies a direct correlation between $T_c$ and NbN crystalline quality [22–24,26]. The $T_c$ values of our NbN films with an AlN buffer layer on 300 mm Si wafer grown at a substrate temperature of 400 °C were comparable with the $T_c$ values of NbN films with a similar thickness and an AlN buffer layer on 200 mm Si wafer [17]. It is noted that the suppression of $T_c$ in thinner NbN films is widely reported, where several physical mechanisms including proximity effect, weak localization, interface, and quantum size effect were proposed to be the possible origins [27–29]. The suppressed $T_c$ at lower film thickness could also be attributed to surface contamination such as carbon and oxygen as well as the formation of NbOx as a secondary phase which reduces the effective NbN thickness [30]. Furthermore, the $T_c$ of NbN films is extremely sensitive to N concentration. Experimental results have shown that the Ar:$N_2$ ratio during the sputtering deposition of nitride films could lead to a change in the physical properties of the nitride films [31,32].

A linear relationship was found between the transition temperature as a function of the inverse of film thickness. To further explore the critical thickness, below which the superconducting properties vanish, we used the Simonin model to fit our data: $T_c(d) = T_{c0}(1 − d_c/d)$, where $d_c$ is the critical thickness, and $T_{c0}$ corresponds to the maximum $T_c$ [33]. In Figure 3, the data for the films with thickness $d$ = 50 nm deviated from the linear fit (red straight lines), while the rest of the data fit well for films with and without AlN buffers. Similar deviations have been observed in other superconducting films [34]. The fitting parameters $d_c$ = 1.29 ± 0.02 nm and $T_{c0}$ = 15.27 ± 0.04 K for the NbN films with an AlN buffer layer, and $d_c$ = 1.24 ± 0.04 nm and $T_{c0}$ = 13.83 ± 0.18 K for the NbN films without an AlN buffer layer were obtained. It is important to note that the maximum transition temperature $T_{c0}$ was improved substantially for NbN films with an AlN buffer layer while critical thickness $d_c$ remained essentially the same. Considering that $d_c = 2a/N(0)V$, where $a$ is the Thomas–Fermi screening length, $N(0)$ is the density of states at the Fermi surface, and $V$ is the electron–phonon interaction potential, we estimated a lower bound for the scattering length $a$ ~ 0.2 nm for the NbN with and without an AlN buffer by taking $N(0)V$ = 0.32 for bulk NbN and also considering the correction due to the reduced film thickness [29]. This value is on the same order of magnitude as a previous report by Polakovic et al. [35].

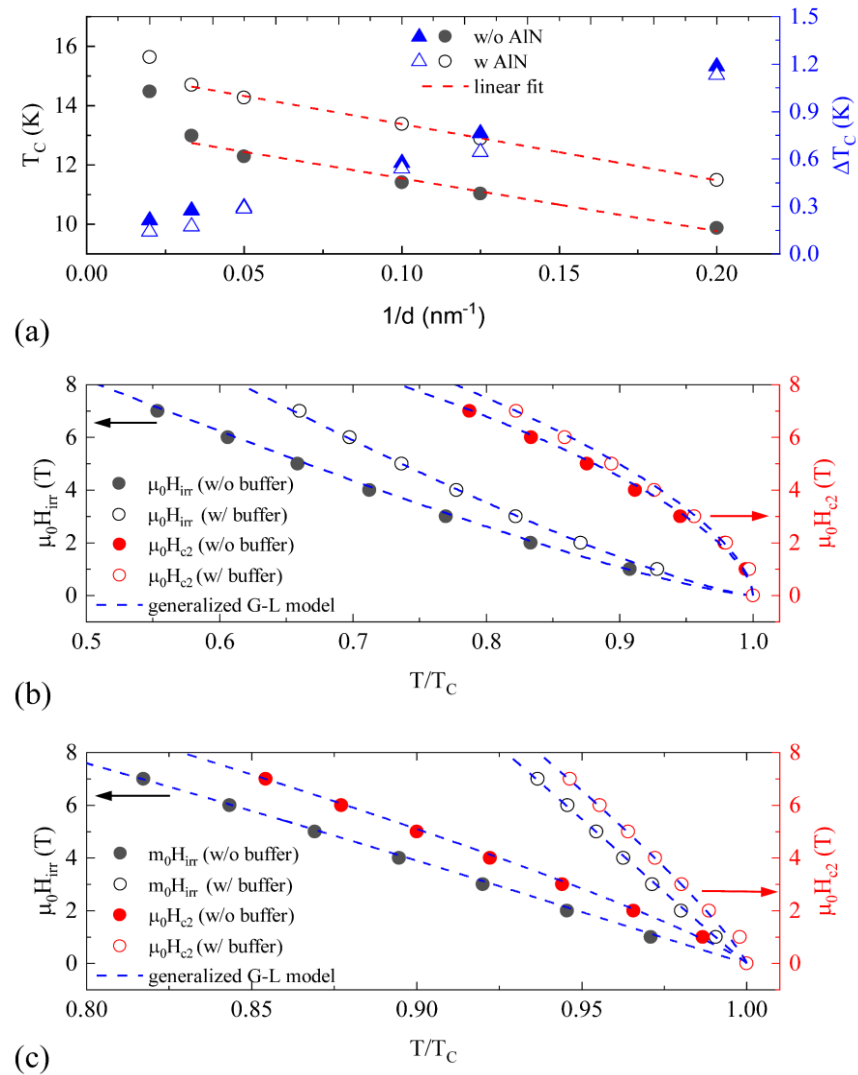

**Figure 4.** (**a**) The transition temperature and the transition width as a function of 1/*d* for NbN films with and without an AlN buffer layer [25]. The upper critical field (right *y*-axis) and the irreversibility field (left *y*-axis) as a function of reduced temperature *T*/*T<sub>C</sub>* for NbN films with the NbN thickness of (**b**) 5 nm and (**c**) 50 nm. The blue-dashed lines represent the generalized G–L model fitting.

Figure 4b,c shows the upper critical field ($H_{c2}$) and irreversibility field ($H_{irr}$) as a function of reduced temperature $T/T_C$ for the 5 nm and 50 nm films, respectively, where the irreversible field $H_{irr}$ is defined as the field of the onset of the transition (10% of the normal resistance) while the upper critical field $H_{c2}$ is defined as the field at the end of the transition (90% of the normal resistance). Our $H_{c2}$ and $H_{irr}$ data can best be fitted by the generalized Ginzburg–Landau theory: $H(T) = H(0) \times (1 - (T/T_C)^a)^b$, where $H(0)$ is the critical field [36]. The fitting parameters *a*, *b*, $H_{c2}$, and $H_{irr}$ at 4.2 K are listed in Table 2. The upper critical field and irreversibility field showed strong thickness dependence, as shown in Table 2. Furthermore, the upper critical field and irreversibility field were both enhanced with an AlN buffer layer for the NbN films at a given thickness. This was especially noticeable for the 50 nm NbN film. The critical field values reported here were higher than those measured on epitaxial NbN grown by MBE and sputtering [37,38]. A possible reason could be the existence of a larger amount of flux pinning sites (grain boundaries between columnar grains as shown in Figure 3) in our NbN films, due to the relatively lower processing temperatures. The detailed physical mechanism of the effect of the AlN buffer layer on the critical field needs to be studied in the future.

**Table 2.** Upper critical field and irreversibility field for 5 nm and 50 nm NbN with and without an AlN buffer layer.

| *d* (nm) | AlN Buffer | $\mu_0 H_{c2}$ (T) | $\mu_0 H_{irr}$ (T) |
|---|---|---|---|
| 5 | Yes | 14.70 ± 0.123 * | 14.00 ± 0.091 * |
| 5 | No | 12.61 ± 0.078 * | 8.89 ± 0.078 * |
| 50 | Yes | 59.19 ± 0.411 ** | 58.41 ± 0.459 ** |
| 50 | No | 25.13 ± 0.242 ** | 20.74 ± 0.223 ** |

* Fitting parameters for 5 nm NbN are *a* = 1.2, *b* = 0.6 for $H_{c2}$ and *a* = 1.2, *b* = 1.3 for $H_{irr}$; ** Fitting parameters for 50 nm NbN are *a* = 1.2, *b* = 0.85 for $H_{c2}$ and *a* = 1.85, *b* = 1.03 for $H_{irr}$.

## 4. Conclusions

In conclusion, we have demonstrated a CMOS BEOL-compatible fabrication process for depositing highly uniform superconducting NbN films on 300 mm Si wafers by an industrial HVM PVD system. The c-axis oriented AlN buffer layer deposited on Si was found to be crucial in achieving highly textured δ-NbN films. Improved superconducting critical temperature, superconducting transition width, and critical fields were observed on NbN films deposited on an AlN buffered Si for NbN film thicknesses ranging from 5 nm to 50 nm. The improved superconducting properties observed in NbN films with an AlN buffer layer could be attributed to the better crystal quality, interface, and homogeneity from the highly textured growth. These results demonstrate the potential of utilizing an industrial HVM thin film deposition system for high-quality and large-scale NbN superconducting material deposition. More importantly, this study paves the way for integrating NbN thin films with other functional materials in a 300 mm advanced CMOS manufacturing environment for the fabrication of large-scale quantum devices.

**Author Contributions:** Conceptualization, Z.Y., N.P., Z.C., H.Z., Q.J. and M.Z.; methodology, Z.Y., H.Z., Q.J. and M.Z.; formal analysis, Z.Y., X.W., P.R., D.Z., P.L., S.D., H.W., N.C., H.Z., Q.J. and M.Z.; investigation, Z.Y., X.W., P.R., D.Z., P.L., S.D., H.W. and N.C.; writing—original draft preparation, Z.Y., P.R., X.W., H.Z., Q.J. and M.Z.; writing—review and editing, Z.Y., N.P., Z.C., H.Z., Q.J. and M.Z.; supervision, Z.Y., Q.J. and M.Z.; All authors have read and agreed to the published version of the manuscript.

**Funding:** The work at the University at Buffalo (UB) was partially supported by both the SUNY Applied Materials Research Institute and the UB's New York State Center of Excellence in Materials Informatics through Co-Funded Projects. D.Z. and H.W. acknowledge the support by the U.S. NSF under DMR-2016453 and DMR-1565822.

**Institutional Review Board Statement:** Not applicabble

**Informed Consent Statement:** Not applicable

**Data Availability Statement:** The data that support the findings of this study are available from the corresponding authors upon reasonable request.

**Acknowledgments:** The work is partially supported by The SUNY Applied Materials Research Institute (SAMRI), a strategic alliance between the State University of New York (SUNY) and Applied Materials, Inc. (Applied). The work at the University at Buffalo (UB) was partially supported by the UB's New York State Center of Excellence in Materials Informatics (CMI) through the Co-Funded Projects between UB faculty and industry collaborators. D.Z. and H.W. acknowledge the support by the U.S. NSF under DMR-2016453 and DMR-1565822. The work at Sandia was performed, in part, at the Center for Integrated Nanotechnologies, an Office of Science User Facility operated for the U.S. Department of Energy (DOE) Office of Science by Los Alamos National Laboratory (contract 89233218NCA000001) and Sandia National Laboratories (contract DE-NA0003525). Sandia National Laboratories is a multiprogram laboratory managed and operated by National Technology and Engineering Solutions of Sandia, LLC, a wholly owned subsidiary of Honeywell International, Inc., for the U.S. Department of Energy's National Nuclear Security Administration under contract DE-NA0003525. This paper describes objective technical results and analysis. Any subjective views or opinions that might be expressed in the paper do not necessarily represent the views of the U.S. Department of Energy or the United States Government.

**Conflicts of Interest:** Authors Zihao Yang, Nag Patibandla, Zhebo Chen and Mingwei Zhu were employed by the company Applied Materials Inc. The remaining authors declare that the research was conducted in the absence of any commercial or financial relationships that could be construed as a potential conflict of interest. The funders had no role in the design of the study; in the collection, analysis, or interpretation of data; in the writing of the manuscript; or in the decision to publish the results.